\pdfoutput=1
\documentclass[12pt,a4paper]{article}
\usepackage{ifthen}
\newboolean{uprightparticles}
\setboolean{uprightparticles}{false} 

\newcommand{\npart}{\ensuremath{\langle \mathrm{N_{part}} \rangle}}
\newcommand{\mpt}{\ensuremath{\langle p_T \rangle}}
\def\sqsnn {\ensuremath{\protect\sqrt{s_{\scriptscriptstyle\text{NN}}}}\xspace}
\def\lhcb   {\mbox{LHCb}\xspace}
\def\geant      {\mbox{\textsc{Geant4}}\xspace}
\def\pythia     {\mbox{\textsc{Pythia}}\xspace}

\def\BF         {{\ensuremath{\mathcal{B}}}\xspace}
\def\BR         {\BF}
\def\Y#1S{\ensuremath{\PUpsilon{(#1S)}}\xspace}
\def\OneS  {{\Y1S}}
\def\TwoS  {{\Y2S}}
\def\ThreeS{{\Y3S}}
\newcommand{\etot}{{\ensuremath{\varepsilon^{i}_{\mathrm{ tot}}}}\xspace}
\newcommand{\nospaceunit}[1]{\ensuremath{\text{#1}}}       
\newcommand{\aunit}[1]{\ensuremath{\text{\,#1}}}       
     
\newcommand{\gev}{\aunit{Ge\kern -0.1em V}\xspace}
\newcommand{\gevc}{\ensuremath{\aunit{Ge\kern -0.1em V\!/}c}\xspace}
\newcommand{\tev}{\aunit{Te\kern -0.1em V}\xspace}
\newcommand{\TeV}{\tev}
\newcommand{\mevc}{\ensuremath{\aunit{Me\kern -0.1em V\!/}c}\xspace}
\def\ps   {\ensuremath{\aunit{ps}}\xspace}

\def\pt         {\ensuremath{p_{\mathrm{T}}}\xspace}
\def\Pb      {\ensuremath{b}\xspace}
\def\Pc      {\ensuremath{c}\xspace}
\def\cquark    {{\ensuremath{\Pc}}\xspace}
\def\bquark    {{\ensuremath{\Pb}}\xspace}

\def\Ppsi        {\ensuremath{\psi}\xspace}    
\def\PJ      {\ensuremath{J}\xspace}  

\def\jpsi     {{\ensuremath{{\PJ\mskip -3mu/\mskip -2mu\Ppsi\mskip 2mu}}}\xspace}
    
\def\PUpsilon    {\ensuremath{\Upsilon}\xspace}  
\def\sPlot{\mbox{\em sPlot}\xspace}
\def\velo   {VELO\xspace}
\def\mum  {\ensuremath{\,\upmu\nospaceunit{m}}\xspace}

\usepackage{ifthen}
\usepackage{makecell}
\newboolean{pdflatex}
\setboolean{pdflatex}{true}

\newboolean{articletitles}
\setboolean{articletitles}{true} 

\def\paperauthors{LHCb collaboration}
\def\paperasciititle{J/psi photo-production in PbPb peripheral collisions at sqrt(sNN)=5 TeV.}
\def\papertitle{\jpsi photo-production\\ in Pb-Pb peripheral collisions\\ at $\sqsnn=5\tev$}
\def\paperkeywords{{High Energy Physics}, {LHCb}} 
\def\papercopyright{\the\year\ CERN for the benefit of the LHCb collaboration} 
\def\paperlicence{CC BY 4.0 licence}
\def\paperlicenceurl{https://creativecommons.org/licenses/by/4.0/}

\usepackage[top=1in, bottom=1.25in, left=1in, right=1in]{geometry}

\usepackage{microtype}
\usepackage{lineno}  
\usepackage{xspace} 
\usepackage{caption} 

\usepackage{graphicx} 
\usepackage{color}
\usepackage{colortbl}

\usepackage{amsmath} 
\usepackage{amssymb}
\usepackage{amsfonts}
\usepackage{upgreek} 

\newcommand*\patchAmsMathEnvironmentForLineno[1]{%
\expandafter\let\csname old#1\expandafter\endcsname\csname #1\endcsname
\expandafter\let\csname oldend#1\expandafter\endcsname\csname
end#1\endcsname
 \renewenvironment{#1}%
   {\linenomath\csname old#1\endcsname}%
   {\csname oldend#1\endcsname\endlinenomath}%
}
\newcommand*\patchBothAmsMathEnvironmentsForLineno[1]{%
  \patchAmsMathEnvironmentForLineno{#1}%
  \patchAmsMathEnvironmentForLineno{#1*}%
}
\AtBeginDocument{%
\patchBothAmsMathEnvironmentsForLineno{equation}%
\patchBothAmsMathEnvironmentsForLineno{align}%
\patchBothAmsMathEnvironmentsForLineno{flalign}%
\patchBothAmsMathEnvironmentsForLineno{alignat}%
\patchBothAmsMathEnvironmentsForLineno{gather}%
\patchBothAmsMathEnvironmentsForLineno{multline}%
\patchBothAmsMathEnvironmentsForLineno{eqnarray}%
}

\usepackage{hyperxmp}

\usepackage[pdftex,
            pdfauthor={\paperauthors},
            pdftitle={\paperasciititle},
            pdfkeywords={\paperkeywords},
            pdfcopyright={Copyright (C) \papercopyright},
            pdflicenseurl={\paperlicenceurl}]{hyperref}

\usepackage[colorinlistoftodos,textsize=scriptsize]{todonotes}

\usepackage[bottom,flushmargin,hang,multiple]{footmisc}

\usepackage[all]{hypcap}
\usepackage{longtable} 
\usepackage{cite} 
\usepackage{mciteplus}

\begin{document}

\renewcommand{\thefootnote}{\fnsymbol{footnote}}
\setcounter{footnote}{1}

\begin{titlepage}
\pagenumbering{roman}

\vspace*{-1.5cm}
\centerline{\large EUROPEAN ORGANIZATION FOR NUCLEAR RESEARCH (CERN)}
\vspace*{1.5cm}
\noindent
\begin{tabular*}{\linewidth}{lc@{\extracolsep{\fill}}r@{\extracolsep{0pt}}}
\ifthenelse{\boolean{pdflatex}}
 & & \\
 & & CERN-EP-2021-122 \\  
 & & LHCb-PAPER-2020-043 \\  
 
 & & March 3, 2022 \\ 
 & & \\
\end{tabular*}

\vspace*{3.0cm}

{\normalfont\bfseries\boldmath\huge
\begin{center}
  \papertitle 
\end{center}
}

\vspace*{2.0cm}

\begin{center}
\paperauthors\footnote{Authors are listed at the end of this Letter.}
\end{center}

\vspace{\fill}

\begin{abstract}
  \noindent
 The photo-production of \jpsi\ mesons at low transverse momentum is studied in peripheral lead-lead collisions collected by the LHCb collaboration at a centre-of-mass energy per nucleon pair of
 $5~\tev$, corresponding to an integrated luminosity of 210~$\mathrm{\mu b}^{-1}$. 
 The \jpsi\ candidates are reconstructed 
 through the prompt decay into two muons of opposite charge
 in the rapidity region of $2.0<y<4.5$. The results significantly improve previous measurements and are compared to the latest theoretical prediction.

\end{abstract}

\vspace*{2.0cm}

\begin{center}
  Submitted to
  Phys.~Rev.~C.~105~(2022)~L032201
\end{center}

\vspace{\fill}

{\footnotesize 
\centerline{\copyright~\papercopyright. \href{\paperlicenceurl}{\paperlicence}.}}
\vspace*{2mm}

\end{titlepage}

\newpage
\setcounter{page}{2}
\mbox{~}


\renewcommand{\thefootnote}{\arabic{footnote}}
\setcounter{footnote}{0}

\cleardoublepage


\pagestyle{plain} 
\setcounter{page}{1}
\pagenumbering{arabic}



\label{sec:introduction}

One of the main open subjects in ultra-relativistic heavy-ion physics
is the study of the quark-gluon plasma (QGP), an exotic
state of hadronic matter predicted by quantum chromodynamics
(QCD). Quantitative predictions of QGP properties are obtained from
lattice computations~\cite{Aoki:2006we}. Experimentally, one of
the signatures of the QGP formation inside heavy-nuclei collisions is the suppression of heavy quarkonia production, such as the \jpsi particle~\cite{Matsui:1986dk}. The suppression is expected to 
depend on both the temperature of the medium and the binding energy of the
state~\cite{Yan:2006ve}. Cold nuclear matter effects~\cite{Ferreiro:2008wc} seem to influence the measurements in nuclei collisions. These effects must be understood prior to providing a sound interpretation in the QGP framework of the hadronically produced (through the interaction of two partons) quarkonia suppression observed at RHIC and the LHC~\cite{Acharya:2019iur,Khachatryan:2016xxp,Chatrchyan:2011pe,Aaboud:2017cif,Sirunyan:2017lzi,Abelev:2014oea}.

The
ALICE~\cite{Adam:2015gba} and
STAR~\cite{PhysRevLett.123.132302} collaborations measured an excess with respect to expectations from purely hadronic production of \jpsi mesons at very low \pt (below $300\mevc$), where \pt is the component of the
 \jpsi momentum transverse to the beam, in hadronic lead-lead (PbPb) collisions at the center-of-mass energy per nucleon pair
$\sqsnn=2.76\tev$ and 
gold-gold (uranium-uranium)
collisions at $\sqsnn=200\gev$ ($193\gev$). It was posited that this excess
is due to photo-produced \jpsi mesons,
caused by the coherent interaction of the 
large electromagnetic fields generated by the projectile
with the target nucleus~\cite{PhysRev.45.729}.
These types of interactions
were primarily expected to only occur in ultra-peripheral collisions
(UPCs)~\cite{BALTZ_2008}, in which the impact
parameter, $b$, is larger than the sum of the radii $R_a$ and $R_b$ of the two colliding
nuclei, hence without nuclear break-up of the target or the projectile.

In hadronic collisions, the nucleus breaks up, so no coherent production was expected.
A precise measurement of the postulated coherent \jpsi\ production in hadronic collisions would shed light on the coherence of the interaction and on the profile of the photon flux in peripheral PbPb collisions~\cite{ducati_heavy_2018,cepila,PhysRevC.97.044910}. 

In this Letter, a measurement of prompt \jpsi production at very-low \pt\ in PbPb collisions at a centre-of-mass
energy per nucleon pair $\sqsnn=5\tev$ 
is reported. For the first time at the LHC, the production yield is measured versus \pt\ and rapidity,  $y$. The data
were recorded by the LHCb detector in 2018
and correspond to an integrated luminosity of about $210~\upmu\mathrm{b}^{-1}$. 

The \lhcb detector~\cite{LHCb-DP-2008-001,LHCb-DP-2014-002} is a single-arm forward
spectrometer covering the \mbox{pseudorapidity} range $2<\eta <5$,
designed for the study of particles containing \bquark or \cquark
quarks. The detector includes a high-precision tracking system
consisting of a silicon-strip vertex detector surrounding the PbPb
interaction region~\cite{LHCb-DP-2014-001}, a large-area silicon-strip detector located
upstream of a dipole magnet with a bending power of about
$4{\mathrm{\,Tm}}$, and three stations of silicon-strip detectors and straw
drift tubes~\cite{LHCb-DP-2017-001}
placed downstream of the magnet.
The tracking system provides a measurement of the momentum of charged particles with
a relative uncertainty that varies from 0.5\% at low momentum to 1.0\% at 200\gevc.
The minimum distance of a track to a primary PbPb collision vertex (PV), the impact parameter (IP), 
is measured with a resolution of $(15+29/\pt)\mum$,
where \pt is the component of the momentum transverse to the beam, in\,\gevc.
Different types of charged hadrons are distinguished using information
from two ring-imaging Cherenkov detectors~\cite{LHCb-DP-2012-003}. 
Photons, electrons and hadrons are identified by a calorimeter system consisting of
scintillating-pad and preshower detectors, an electromagnetic
and a hadronic calorimeter. Muons are identified by a
system composed of alternating layers of iron and multiwire
proportional chambers~\cite{LHCb-DP-2012-002}.

The trigger consists of a hardware stage, using
information from the calorimeter and muon systems, followed by a software stage,
which applies selections on the fully reconstructed event.

At the hardware trigger stage, events are required to have a muon with high \pt or a
hadron, photon or electron with high transverse energy in the calorimeters.
Two samples are selected for this analysis, a signal sample and a sample of minimum bias events used to normalise through the total number of inelastic events. 
 They have different trigger strategies. Signal events are selected if the number of clusters in the \velo\, $N_{\rm c}$, is  $6000<N_{\rm c}<10000$ or if  $N_{\rm c}<6000$ and two muons with $\pt > 400\mevc$ are reconstructed.

The hardware trigger for the minimum bias (MB) events requires a minimum energy deposit in one of the sub detectors (VELO, calorimeters, muon chambers), and only events with $N_{\rm c}<$ 10000 are kept.
To improve the signal purity, events passing the software trigger are selected if the muon candidates have a \mbox{\pt $>$ 700\mevc},
the particles are consistent with originating from a primary PbPb collision vertex (PV) and are
identified as muons. 
The prompt \jpsi\ candidates, which include feed-down from excited charmonium states originating from $b$-hadron decays, are separated from the non-prompt candidates using the requirement $t_z<0.3\ps$, where $t_z$ is the pseudo decay-time defined as $(z_\jpsi-z_{\rm PV}) m_{\jpsi}/p_z$. Here, $(z_{\jpsi}-z_{\rm PV})$ and $p_z$ are the distance between the \jpsi candidate decay vertex and the PV, and the candidate momentum along the beam axis, respectively, and
$m_\jpsi$ is the known \jpsi\ mass~\cite{10.1093/ptep/ptaa104}.

During data taking, neon (Ne) gas was injected into the beam pipe near
the interaction point using LHCb's SMOG system~\cite{LHCb-PAPER-2014-047} to record fixed-target collisions simultaneously with the PbPb collisions.
These fixed-target PbNe collisions are rejected by placing requirements on the position of the PV.
In order to remove potential contamination from UPCs, which may bias the measurement especially for very low event activity, a minimal
energy deposit in the electromagnetic calorimeter (ECAL) is also required ($E_{\rm tot}>585$ GeV).

\begin{figure}[!tbp]
 \centering{
       \includegraphics[width=0.5\textwidth]{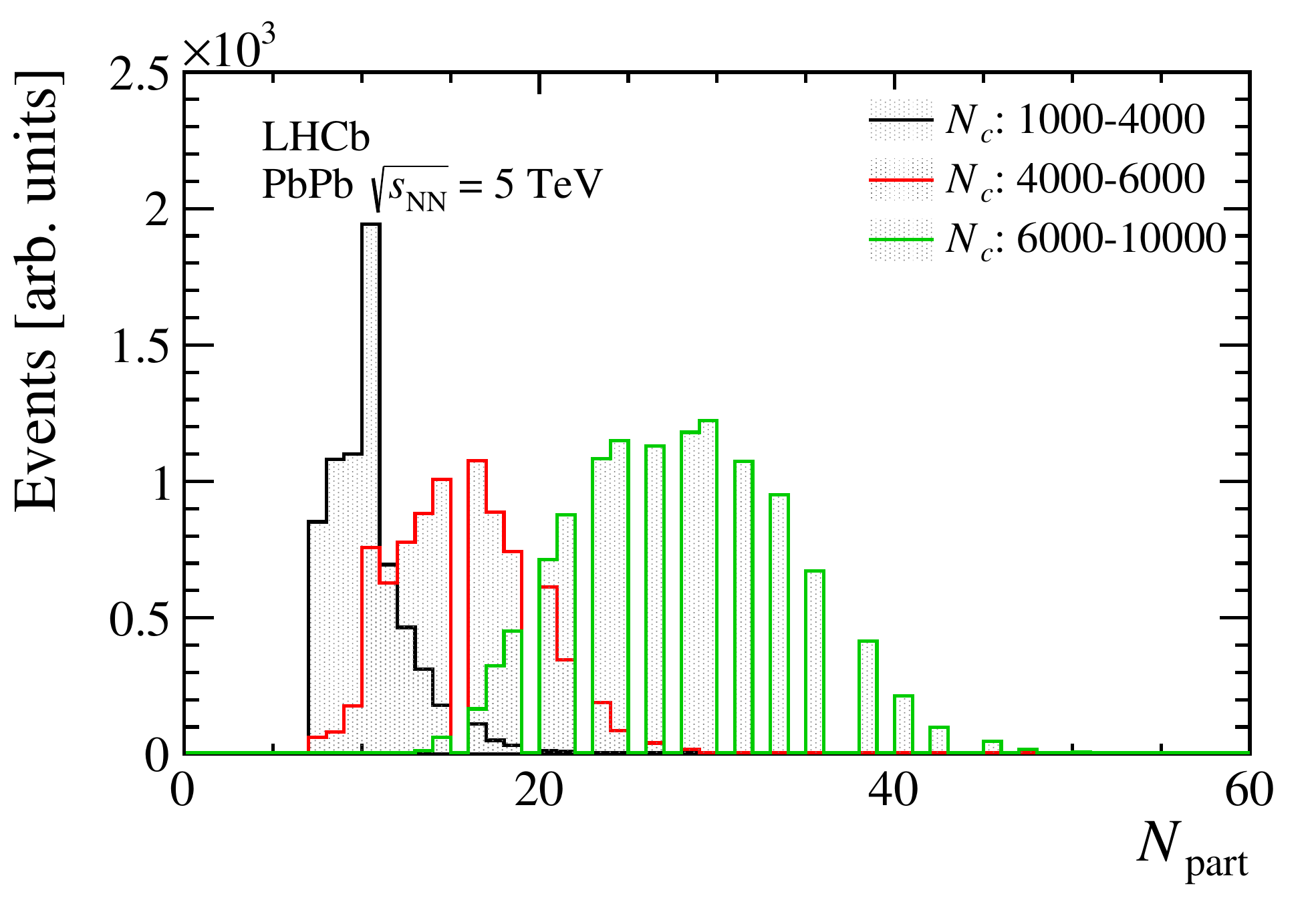}
     \caption{Distribution of $N_\textrm{part}$ in the three $N_c$ intervals of the minimum bias events defined in Table~\ref{tab:centclassvelo}\label{fig:npartdistri}.}}
   \end{figure}

The PbPb sample is divided into intervals of $N_{\rm c}$ which correspond to different numbers of participating nucleons, $N_\textrm{part}$. This quantity is related to the centrality, defined as the percentile of the total inelastic hadronic PbPb cross-section as a function of the released collision energy, which can be approximated by the total energy deposit in ECAL ($E_{\rm tot}$). The more central is the collision, the larger $E_{\rm tot}$ is and the larger $N_\textrm{part}$ is. The percentiles are determined using the
Glauber Monte Carlo (GMC) model~\cite{PhysRevC.97.054910,Abelev:2013qoq}. The model is used to perform a binned fit to the $E_{\rm tot}$ distribution of the MB data sample, collected with the same detector
conditions as the signal sample.
The quantity $N_\textrm{part}$ is estimated for each collision and the mean value, \npart, is derived from events within a given $N_c$ range. Results for each $N_c$ interval are summarized in Table~\ref{tab:centclassvelo}. The distribution of $N_\textrm{part}$ in the three $N_c$ intervals is shown in Fig.~\ref{fig:npartdistri}. More details on the centrality determination in LHCb can be found in Ref.~\cite{LHCb-DP-2021-001} and in the supplemental material \cite{supp_mat}.

  \begin{table}[!tbp]
    \centering
    \caption{Average number of participant nucleons, \npart, along with the corresponding standard deviation of the $N_\textrm{part}$ distribution, $\sigma_{\rm part}$, for each of the selected $N_c$ intervals.}
    \begin{tabular}{ccc}
      $N_c$ & \npart\  & $\sigma_{\rm part}$\\ 
      \hline
      \phantom{0}1000~\textendash~4000\phantom{00}  & 10.6 & 2.9  \\ 
      4000~\textendash~6000\phantom{0}  & 15.7 & 4.1  \\ 
      6000~\textendash~10000 & 27.8 & 7.2\\ 
      1000~\textendash~10000 & 19.7 & 9.2\\
      \hline
    \end{tabular}
    \label{tab:centclassvelo}
  \end{table}

\begin{figure}[!tbp]
 \centering{
       \begin{tabular}{cc}
       \includegraphics[width=0.5\textwidth]{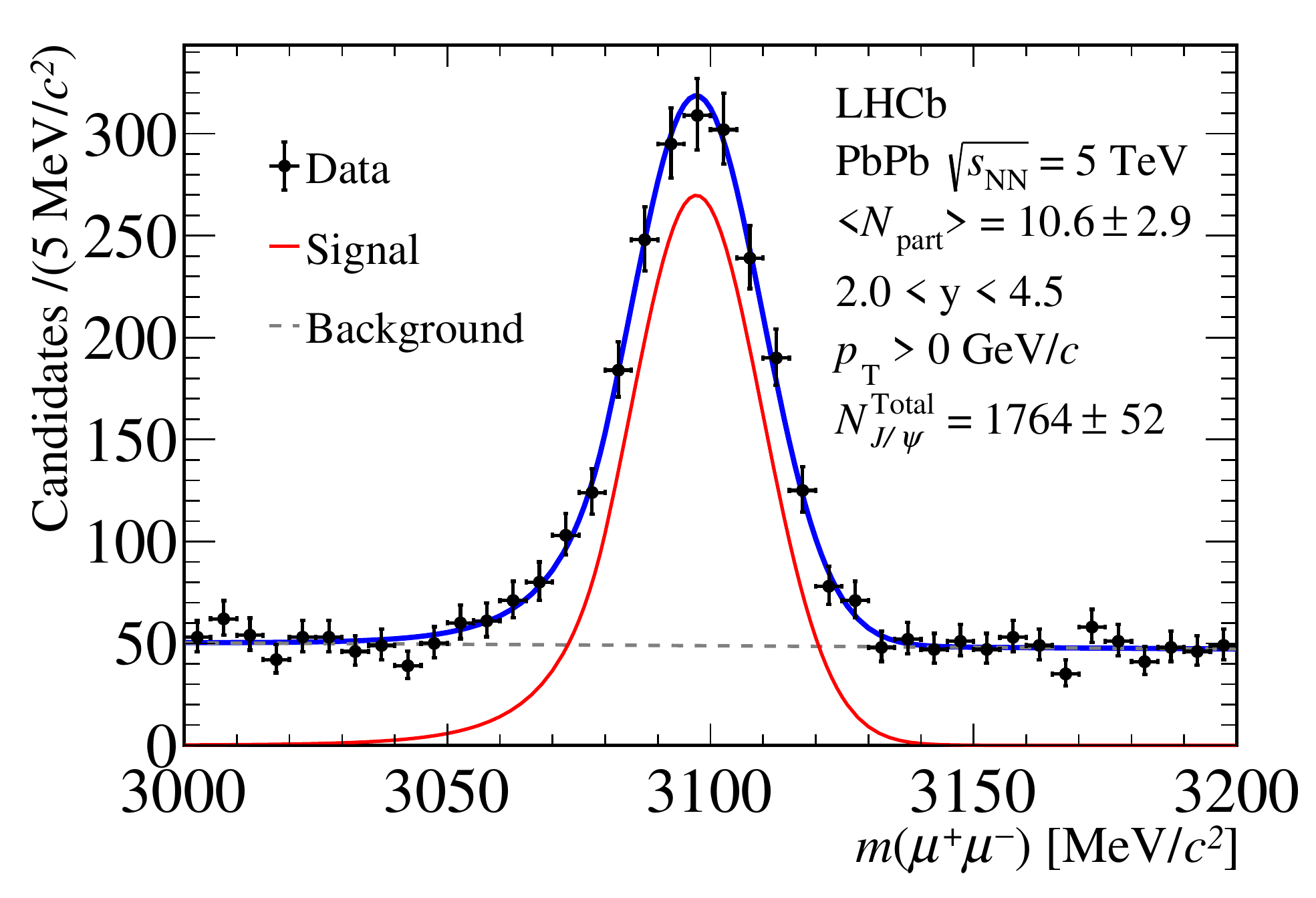} &
        \includegraphics[width=0.5\textwidth]{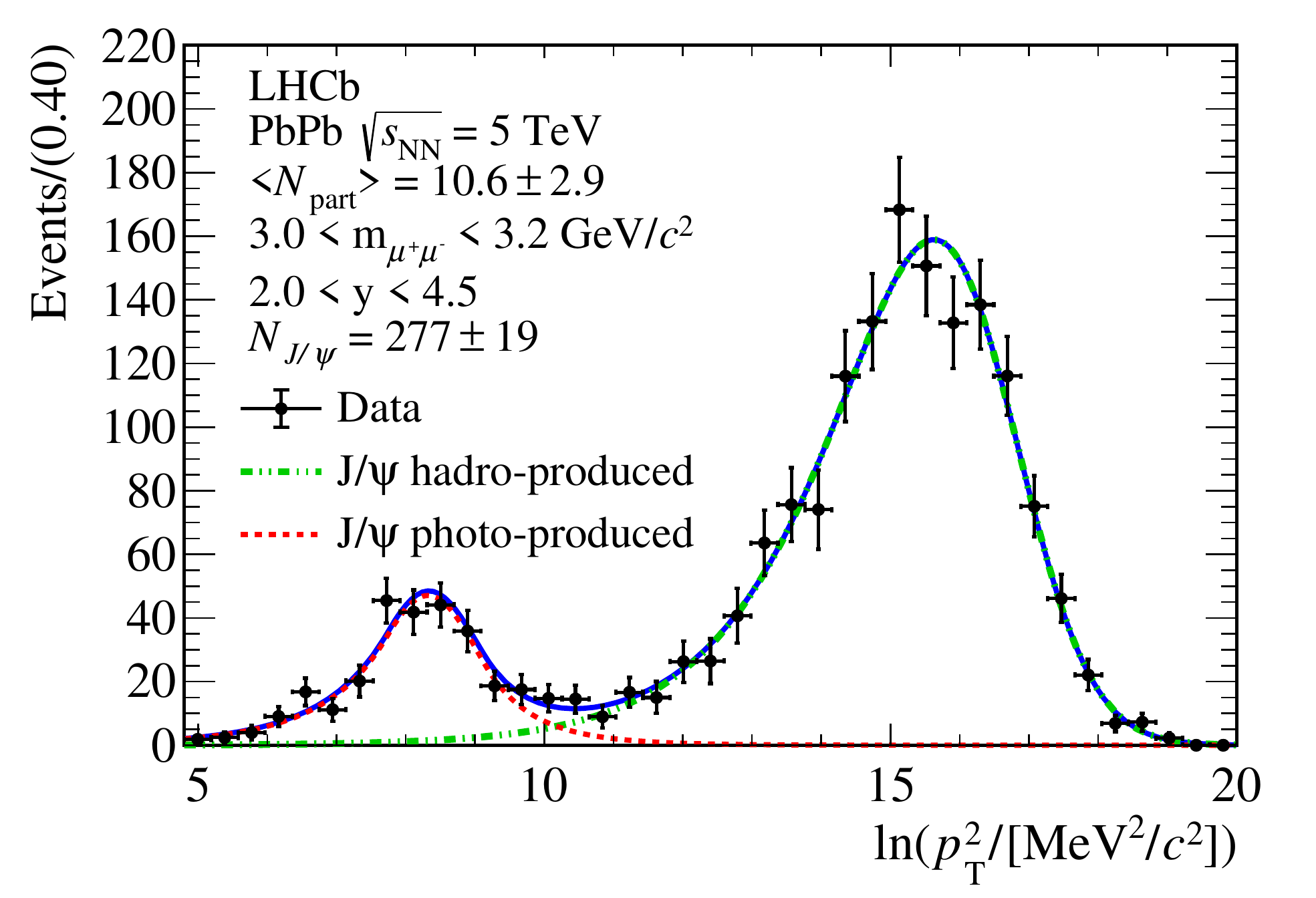}\\
        \end{tabular}
     \caption{(Left) Invariant mass distribution of \jpsi candidates with
      $\npart=10.6\pm2.9$, for $\pt<15.0\gevc$ and $2.0<y <4.5$.
      (Right) Distribution of $\ln(\pt^2)$ of the \jpsi\ candidates for $\npart=10.6\pm2.9$ after background subtraction. The projections of the fit
    to disentangle the coherently photo-produced and hadronically produced \jpsi mesons are overlaid.\label{fig:InvariantMassCen}}}
   \end{figure}

In this Letter, the \jpsi photo-production differential yield is
measured, defined as
  \begin{equation}\label{eq:yields}
    \frac{\mathrm{d}Y^{i}_{\jpsi}}{\mathrm{d}y}  = \frac{N^i_\jpsi}{\BR\, N^i_{\mathrm{MB}}\, \etot\,\Delta y},
  \end{equation}
  
  \begin{equation}\label{eq:yields2}
      \frac{\mathrm{d}^{2}Y^{i}_{\jpsi}}{\mathrm{d}\pt dy} = \frac{\mathrm{d}Y^{i}_{\jpsi}}{\mathrm{d}y} \frac{1}{\Delta\pt},
  \end{equation}
  
where 
$i$ indicates the $N_c$ range, ${N^i_\jpsi}$ is the number of photo-produced \jpsi meson candidates reconstructed through the \mbox{$\jpsi\to\mu^{+} \mu^{-}$} decay channel in the $(\pt,y)$ interval of width $(\Delta \pt,\Delta y)$,
$\BR=(5.961\pm0.033)\%$~\cite{10.1093/ptep/ptaa104} is the branching
     fraction of the decay $\jpsi\to\mu^{+} \mu^{-}$, $N^i_{\mathrm{MB}}$ is the total number of MB
     events, and
     $\etot$ 
     is the efficiency to reconstruct and select the \jpsi\ candidates.
  The dimuon invariant mass, $m(\mu^+\mu^-)$, of the selected candidates is shown in Fig.~\ref{fig:InvariantMassCen} for a representative centrality interval
  for \jpsi candidates with \mbox{\pt$<$15.0\gevc} and $2.0<y<4.5$. An unbinned fit of these candidates is performed using a Crystal-Ball (CB) function for the signal and a first order polynomial for the background.

  Photo-produced \jpsi\ mesons
  and hadronically produced \jpsi\ mesons are then
  disentangled  through an unbinned maximum likelihood fit to the
  dimuon \pt\ spectrum. The fit is performed after subtracting correlated background, i.e. arising from $\gamma \gamma \rightarrow \mu^+\mu^-$ or the Drell-Yan process, and combinatorial background from uncorrelated muon pairs which is the dominant source, using the \sPlot method~\cite{Pivk:2004ty} with $m(\mu^+\mu^-)$ as discriminating variable. To cross check the validity of the \sPlot\  method, the kinematic distributions (\pt\, $y$, $N_{c}$) of the estimated background are compared to the same (normalised) distributions in two invariant mass ranges below and above the resonance peak. A very good agreement is found. The empirical fit model comprises a double-sided Crystal-Ball 
  function~\cite{Skwarnicki:1986xj} expressed in ln($\pt^2$) for the photo-production contribution and a function for the hadronic component that typically has a larger \pt\,
  \begin{equation}
             f(\pt) = \frac{\pt^{n_{1}}}{ \Big[1 + (\frac{\pt}{p_{0}})^{n_{2}}\Big]^{n_{3}}},
	\label{eq:function}
 \end{equation}
  where $n_{1}$, $n_{2}$, $n_{3}$ and $p_{0}$ are parameters free to vary in the fit.
  The projections of the fits in the centrality interval \npart=$10.6\pm 2.9$ are shown in Fig.~\ref{fig:InvariantMassCen}, overlaid to the data distributions. A good description of the data is observed in all centrality intervals. The photo-produced \jpsi candidates are visible in the range $0 < \pt < 250\mevc$. The \pt\ distribution of the photo-produced \jpsi candidates does not rise towards vanishing $\pt$ due to the interference caused by the negative parity of the photon as explained in Ref.~\cite{PhysRevC.97.044910}. 

  Simulation is required to model the effects of the detector acceptance and of the
  selection requirements on the signal.
  The PbPb collisions are generated using EPOS~\cite{EPOS} and the hard
  process is generated with \pythia~\cite{Sjostrand:2007gs,*Sjostrand:2006za}
  with a specific \lhcb configuration~\cite{LHCb-PROC-2010-056}.
  
  An additional signal sample where the \jpsi is transversely polarised
was produced using the STARlight~\cite{starlight} generator to study the acceptance assuming the coherent photo-production scenario.
  The interactions of the generated particles with the detector, and its response,
  are implemented using the \geant
  toolkit~\cite{Allison:2006ve, *Agostinelli:2002hh} as described in
  Ref.~\cite{LHCb-PROC-2011-006}.
  The total efficiency is determined independently in each interval
  of centrality, and it includes the effects of the geometrical
  acceptance ($\epsilon_{\rm acc}$), the trigger efficiency ($\epsilon_{\rm trigger}$), the reconstruction and selection efficiency
  ($\epsilon_{\rm rec\&sel}$),
  and the efficiency of the particle identification (PID) criteria ($\epsilon_{\rm PID}$).
  The acceptance is determined using the STARlight sample in the kinematic range of the analysis.
  The efficiency $\epsilon_{\rm rec\&sel}$ is estimated using
  simulation and data calibration techniques. The main component of the
  reconstruction inefficiency is due to the tracking algorithms,
  as the performance is affected by the high occupancy in PbPb collisions.
  The relative reconstruction efficiency between data and simulation is evaluated
  using two $D^0$ meson decay channels ($D^{0}\rightarrow K^{-}\pi^{+}$ and $D^{0} \rightarrow K^{-}\pi^{+}\pi^{-}\pi^{+}$). 
 The yields are evaluated in PbPb data and simulation
and the difference of their ratio to unity is encoded in a factor $k(N_{c})$.
  This factor depends on the event multiplicity with $k$ ranging from 0.97 to 0.91 with increasing $N_c$, assuming $k(N_{c})$ is the same for the $\pi$ and 
  $\mu$ tracks. The latter factor is used to correct the reconstructed \jpsi\ candidates in simulation. An additional correction is applied to correct discrepancies between reconstructed \jpsi\ kinematic distributions by weighting the variables $N_{c}$, \pt, and $y$ of the \jpsi in simulation to match the data.
  The PID efficiency $\epsilon_{\rm PID}$
  is evaluated
  using a tag-and-probe approach with $\jpsi \rightarrow \mu\mu$ decays reconstructed in proton-proton collisions that provides PID efficiency tables for single muons. 
  Those efficiencies are used to perform a two-dimensional ($p_{T}$, $N_{c}$) extrapolation, using first- and second-order polynomial functions,
  to estimate the decrease of the efficiencies for higher multiplicities seen in PbPb collisions. No extrapolation is performed based on the rapidity as no correlation is seen between $N_{c}$
  and $y$.

  Several sources of systematic uncertainties are considered. The uncertainty associated with the fit model used to evaluate the signal yields is determined by testing alternative fit functions.
The \pt of the hadronically produced \jpsi\ candidates is modelled by a Tsallis function~\cite{tsallis_nonextensive_1999}.
The background shape is also modified to account for incoherent photo-produced \jpsi , defined as the interaction between one photon and a single nucleon implying the destruction of the nucleus.
  This contribution typically produces \jpsi mesons at
  higher \pt than the coherent photo-production source. Therefore,
  another double CB
  function is added to model this potential contribution. The incoherent contribution shares the shape parameters of the coherent contribution with the mean \pt\ and width shifted according to the differences obtained in the STARlight simulations. By computing the difference to the reference fit, a total uncertainty of about 1.3\% averaged over all centrality intervals is obtained.

  The systematic uncertainty associated with the evaluation of the efficiencies is
  divided into uncertainties due to the
  $\epsilon_{\rm rec\&sel}$, $\epsilon_{\rm PID}$, and $\epsilon_{\rm trigger}$ efficiencies.  
  Three systematic effects are considered for the measurement of $\epsilon_{\rm rec\&sel}$:
  the uncertainty on the weighting procedure, the uncertainty associated
  with the evaluation of the factor $k$, and the correlation between the variables \pt\ and $y$.
  The uncertainty on the weighting procedure is estimated by comparing \pt, $y$, and $N_{c}$ of the weighted
  distributions of the \jpsi mesons in simulation with those in data
  after background subtraction. The difference between the two leads to a global uncertainty of 2\%.
  The uncertainty on the factor $k$ is evaluated by varying its value within its uncertainty and propagating it to the tracking efficiency. An uncertainty from 2.9\% to 7.4\% is found depending on the considered multiplicity interval.
  The uncertainty on the correlation between the variables \pt and $y$ is estimated to be 1\% using calibration samples from proton-proton collisions.

  The uncertainty coming from the muon PID efficiency tables is evaluated with a smearing technique. The \jpsi\ PID efficiency is computed using efficiency PID tables with the values in each interval varied within their uncertainty. This procedure is repeated several times and the largest difference is taken as systematic uncertainty; the effect is smaller
  than 1\% and considered negligible compared to the uncertainty given by the difference of the two functions used for the extrapolation.

  The uncertainty on $\epsilon_{\rm trigger}$ is estimated by comparing the trigger efficiency measurement with another method based on data.
  The method consists of evaluating the
  efficiency using a sample selected by the same trigger algorithms
  but independent from those used in the selection of the signal.
  The difference of 3\% between the two methods is taken as systematic
  uncertainty on the trigger efficiency.
   The total systematic uncertainties are obtained by summing in quadrature the different sources of uncertainties.

\begin{figure}[tpb]
        \centering
        \begin{tabular}{cc}
        \includegraphics[width=0.50\textwidth]{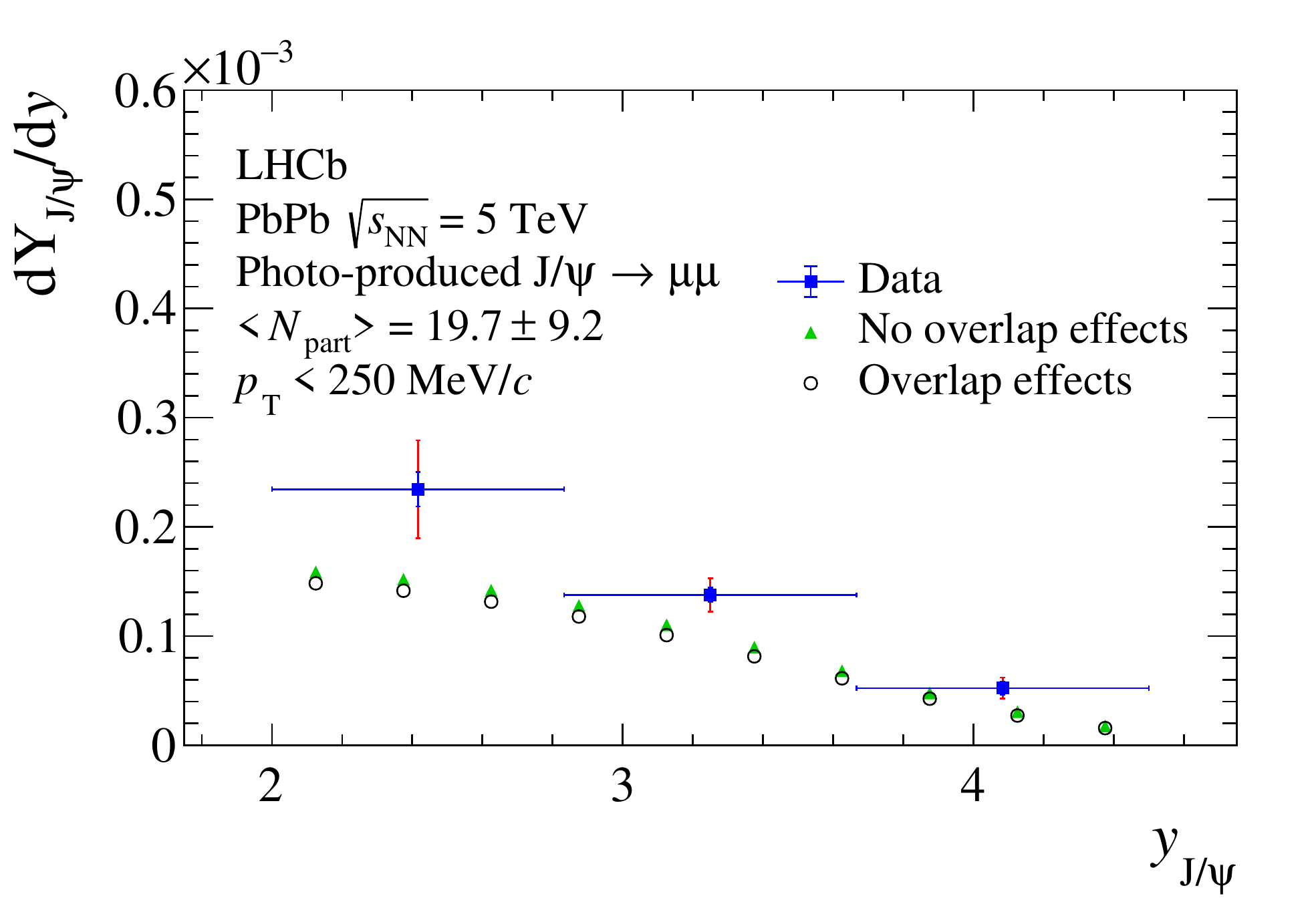}
        
        &
        \includegraphics[width=0.50\textwidth]{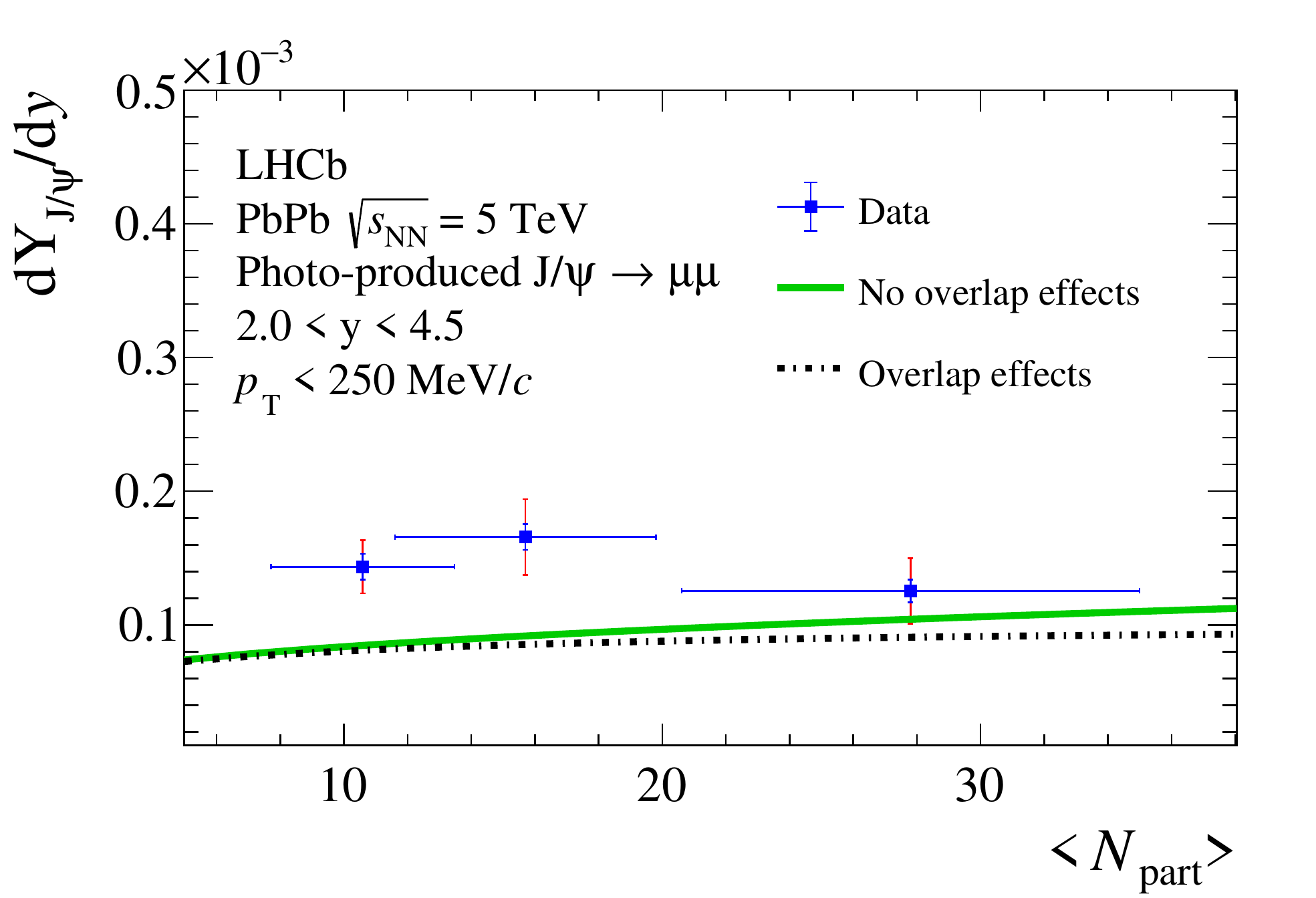}\\
        \end{tabular}\\
        \centering
        \includegraphics[width=0.50\textwidth]{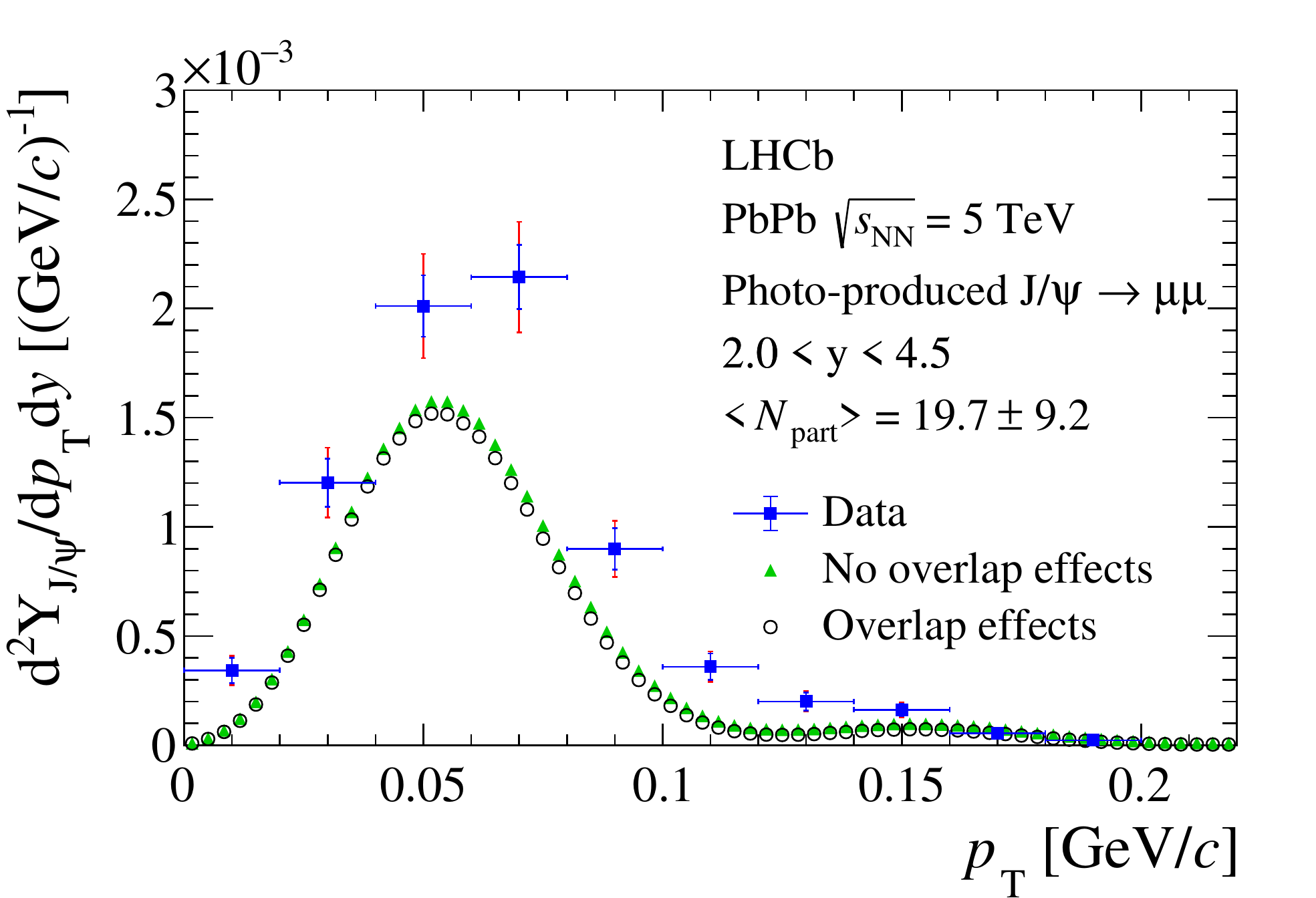}\\
            \caption{(Top left) Differential yields of photo-produced \jpsi\ candidates as a function of rapidity and (top right) \npart. (Bottom) Double differential yields as a function of \pt. The vertical inner blue bars represent the statistical uncertainty and the outer red bars the total uncertainty. The horizontal bars in the \npart\ results correspond to the standard deviation of the $\rm{N_{part}}$ distribution of the MB events. The yields are compared to the prediction from Ref.~\cite{PhysRevC.97.044910,PhysRevC.99.061901} that take (black) or not take (green) into account the effect from the overlap region of the collision.} 
        \label{fig:resultintnpart}
        \end{figure}

 The differential \jpsi\ photo-production yields, Eq.~(\ref{eq:yields}), as a function of  the rapidity for \npart=$19.7\pm 9.2$ and as a function of \npart\  are shown in Fig.~\ref{fig:resultintnpart} (top). The 
double-differential \jpsi\ photo-production yields, Eq.~(\ref{eq:yields2}), as a function of  the transverse momentum are as well shown in Fig.~\ref{fig:resultintnpart} (bottom). The mean \pt\ of the coherent \jpsi is found to be $\mpt$ = 64.9 $\pm$ 2.4 \mevc. The results are compared to the theoretical prediction~\cite{PhysRevC.97.044910,PhysRevC.99.061901} and detailed results are presented in the supplemental material~\cite{supp_mat}. The model assumes two scenarios in which the coherence of the 
\jpsi\ production is (overlap effect) or is not (no overlap effect) affected by interactions with the overlap region of the two colliding nuclei.
Little difference is observed between the two theory scenarios. The divergence is important in more central collisions due to the increase of the
photon flux with a decrease of the impact parameter. In the overlapping scenario, this effect is balanced by excluding the overlapping region from the interaction.

 The measured yield of the coherent \jpsi\ production is higher at low rapidity than at high rapidity and it is consistent with being constant  with respect to \npart\ for the region considered in the analysis.

In summary, the yield of coherently photo-produced
prompt \jpsi mesons at very low \pt\ in peripheral PbPb collisions
collected at \sqsnn= 5 \tev is measured with the LHCb experiment. The yields are studied as a function of rapidity and transverse momentum of the \jpsi meson in intervals
of the number of participant nucleons \npart. These results are the most precise to date, and support the hypothesis of coherent \jpsi\ photo-production in peripheral hadronic collisions
suggested by the other
experiments~\cite{Adam:2015gba,PhysRevLett.123.132302}. 
The shape of the results is qualitatively described by the
theoretical prediction~\cite{PhysRevC.97.044910,PhysRevC.99.061901}, although a normalization discrepancy
is observed which could hide a possible additional contribution.

\section*{Acknowledgements}
%
%
\noindent We express our gratitude to our colleagues in the CERN
accelerator departments for the excellent performance of the LHC. We
thank the technical and administrative staff at the LHCb
institutes.
We acknowledge support from CERN and from the national agencies:
CAPES, CNPq, FAPERJ and FINEP (Brazil); 
MOST and NSFC (China); 
CNRS/IN2P3 (France); 
BMBF, DFG and MPG (Germany); 
INFN (Italy); 
NWO (Netherlands); 
MNiSW and NCN (Poland); 
MEN/IFA (Romania); 
MSHE (Russia); 
MICINN (Spain); 
SNSF and SER (Switzerland); 
NASU (Ukraine); 
STFC (United Kingdom); 
DOE NP and NSF (USA).
We acknowledge the computing resources that are provided by CERN, IN2P3
(France), KIT and DESY (Germany), INFN (Italy), SURF (Netherlands),
PIC (Spain), GridPP (United Kingdom), RRCKI and Yandex
LLC (Russia), CSCS (Switzerland), IFIN-HH (Romania), CBPF (Brazil),
PL-GRID (Poland) and NERSC (USA).
We are indebted to the communities behind the multiple open-source
software packages on which we depend.
Individual groups or members have received support from
ARC and ARDC (Australia);
AvH Foundation (Germany);
EPLANET, Marie Sk\l{}odowska-Curie Actions and ERC (European Union);
A*MIDEX, ANR, IPhU and Labex P2IO, and R\'{e}gion Auvergne-Rh\^{o}ne-Alpes (France);
Key Research Program of Frontier Sciences of CAS, CAS PIFI, CAS CCEPP, 
Fundamental Research Funds for the Central Universities, 
and Sci. \& Tech. Program of Guangzhou (China);
RFBR, RSF and Yandex LLC (Russia);
GVA, XuntaGal and GENCAT (Spain);
the Leverhulme Trust, the Royal Society
 and UKRI (United Kingdom).

\addcontentsline{toc}{section}{References}

\begin{mcitethebibliography}{10}
\mciteSetBstSublistMode{n}
\mciteSetBstMaxWidthForm{subitem}{\alph{mcitesubitemcount})}
\mciteSetBstSublistLabelBeginEnd{\mcitemaxwidthsubitemform\space}
{\relax}{\relax}

\bibitem{Aoki:2006we}
Y.~Aoki {\em et~al.}, \ifthenelse{\boolean{articletitles}}{\emph{{The Order of
  the quantum chromodynamics transition predicted by the standard model of
  particle physics}}, }{}\href{https://doi.org/10.1038/nature05120}{Nature
  \textbf{443} (2006) 675},
  \href{http://arxiv.org/abs/hep-lat/0611014}{{\normalfont\ttfamily
  arXiv:hep-lat/0611014}}\relax
\mciteBstWouldAddEndPuncttrue
\mciteSetBstMidEndSepPunct{\mcitedefaultmidpunct}
{\mcitedefaultendpunct}{\mcitedefaultseppunct}\relax
\EndOfBibitem
\bibitem{Matsui:1986dk}
T.~Matsui and H.~Satz, \ifthenelse{\boolean{articletitles}}{\emph{{\jpsi\
  suppression by quark-gluon plasma formation}},
  }{}\href{https://doi.org/10.1016/0370-2693(86)91404-8}{Phys.\ Lett.\
  \textbf{B178} (1986) 416}\relax
\mciteBstWouldAddEndPuncttrue
\mciteSetBstMidEndSepPunct{\mcitedefaultmidpunct}
{\mcitedefaultendpunct}{\mcitedefaultseppunct}\relax
\EndOfBibitem
\bibitem{Yan:2006ve}
L.~Yan, P.~Zhuang, and N.~Xu,
  \ifthenelse{\boolean{articletitles}}{\emph{{Competition between $J/\psi$
  suppression and regeneration in quark-gluon plasma}},
  }{}\href{https://doi.org/10.1103/PhysRevLett.97.232301}{Phys.\ Rev.\ Lett.\
  \textbf{97} (2006) 232301},
  \href{http://arxiv.org/abs/nucl-th/0608010}{{\normalfont\ttfamily
  arXiv:nucl-th/0608010}}\relax
\mciteBstWouldAddEndPuncttrue
\mciteSetBstMidEndSepPunct{\mcitedefaultmidpunct}
{\mcitedefaultendpunct}{\mcitedefaultseppunct}\relax
\EndOfBibitem
\bibitem{Ferreiro:2008wc}
E.~G. Ferreiro, F.~Fleuret, J.~P. Lansberg, and A.~Rakotozafindrabe,
  \ifthenelse{\boolean{articletitles}}{\emph{{Cold nuclear matter effects on
  $J/\psi$ production: Intrinsic and extrinsic transverse momentum effects}},
  }{}\href{https://doi.org/10.1016/j.physletb.2009.07.076}{Phys.\ Lett.\
  \textbf{B680} (2009) 50},
  \href{http://arxiv.org/abs/0809.4684}{{\normalfont\ttfamily
  arXiv:0809.4684}}\relax
\mciteBstWouldAddEndPuncttrue
\mciteSetBstMidEndSepPunct{\mcitedefaultmidpunct}
{\mcitedefaultendpunct}{\mcitedefaultseppunct}\relax
\EndOfBibitem
\bibitem{Acharya:2019iur}
ALICE collaboration, S.~Acharya {\em et~al.},
  \ifthenelse{\boolean{articletitles}}{\emph{{Studies of J/$\psi$ production at
  forward rapidity in Pb-Pb collisions at $\sqrt{s_{\rm{NN}}}$ = 5.02 TeV}},
  }{}\href{https://doi.org/10.1007/JHEP02(2020)041}{JHEP \textbf{02} (2020)
  041}, \href{http://arxiv.org/abs/1909.03158}{{\normalfont\ttfamily
  arXiv:1909.03158}}\relax
\mciteBstWouldAddEndPuncttrue
\mciteSetBstMidEndSepPunct{\mcitedefaultmidpunct}
{\mcitedefaultendpunct}{\mcitedefaultseppunct}\relax
\EndOfBibitem
\bibitem{Khachatryan:2016xxp}
CMS collaboration, V.~Khachatryan {\em et~al.},
  \ifthenelse{\boolean{articletitles}}{\emph{{Suppression of $\OneS, \TwoS$ and
  $\ThreeS$ quarkonium states in PbPb collisions at \sqsnn = 2.76 TeV}},
  }{}\href{https://doi.org/10.1016/j.physletb.2017.04.031}{Phys.\ Lett.\
  \textbf{B770} (2017) 357},
  \href{http://arxiv.org/abs/1611.01510}{{\normalfont\ttfamily
  arXiv:1611.01510}}\relax
\mciteBstWouldAddEndPuncttrue
\mciteSetBstMidEndSepPunct{\mcitedefaultmidpunct}
{\mcitedefaultendpunct}{\mcitedefaultseppunct}\relax
\EndOfBibitem
\bibitem{Chatrchyan:2011pe}
CMS collaboration, S.~Chatrchyan {\em et~al.},
  \ifthenelse{\boolean{articletitles}}{\emph{{Indications of suppression of
  excited $\PUpsilon$ states in PbPb collisions at \sqsnn = 2.76 TeV}},
  }{}\href{https://doi.org/10.1103/PhysRevLett.107.052302}{Phys.\ Rev.\ Lett.\
  \textbf{107} (2011) 052302},
  \href{http://arxiv.org/abs/1105.4894}{{\normalfont\ttfamily
  arXiv:1105.4894}}\relax
\mciteBstWouldAddEndPuncttrue
\mciteSetBstMidEndSepPunct{\mcitedefaultmidpunct}
{\mcitedefaultendpunct}{\mcitedefaultseppunct}\relax
\EndOfBibitem
\bibitem{Aaboud:2017cif}
ATLAS collaboration, M.~Aaboud {\em et~al.},
  \ifthenelse{\boolean{articletitles}}{\emph{{Measurement of quarkonium
  production in proton–lead and proton–proton collisions at $5.02\tev$ with
  the ATLAS detector}},
  }{}\href{https://doi.org/10.1140/epjc/s10052-018-5624-4}{Eur.\ Phys.\ J.\
  \textbf{C78} (2018) 171},
  \href{http://arxiv.org/abs/1709.03089}{{\normalfont\ttfamily
  arXiv:1709.03089}}\relax
\mciteBstWouldAddEndPuncttrue
\mciteSetBstMidEndSepPunct{\mcitedefaultmidpunct}
{\mcitedefaultendpunct}{\mcitedefaultseppunct}\relax
\EndOfBibitem
\bibitem{Sirunyan:2017lzi}
CMS collaboration, A.~M. Sirunyan {\em et~al.},
  \ifthenelse{\boolean{articletitles}}{\emph{{Suppression of excited
  $\PUpsilon$ states relative to the ground state in Pb-Pb collisions at \sqsnn
  = 5.02 \TeV}},
  }{}\href{https://doi.org/10.1103/PhysRevLett.120.142301}{Phys.\ Rev.\ Lett.\
  \textbf{120} (2018) 142301},
  \href{http://arxiv.org/abs/1706.05984}{{\normalfont\ttfamily
  arXiv:1706.05984}}\relax
\mciteBstWouldAddEndPuncttrue
\mciteSetBstMidEndSepPunct{\mcitedefaultmidpunct}
{\mcitedefaultendpunct}{\mcitedefaultseppunct}\relax
\EndOfBibitem
\bibitem{Abelev:2014oea}
ALICE collaboration, B.~B. Abelev {\em et~al.},
  \ifthenelse{\boolean{articletitles}}{\emph{{Production of inclusive
  $\PUpsilon$(1S) and $\PUpsilon$(2S) in p-Pb collisions at $\sqsnn = 5.02
  \TeV$}}, }{}\href{https://doi.org/10.1016/j.physletb.2014.11.041}{Phys.\
  Lett.\  \textbf{B740} (2015) 105},
  \href{http://arxiv.org/abs/1410.2234}{{\normalfont\ttfamily
  arXiv:1410.2234}}\relax
\mciteBstWouldAddEndPuncttrue
\mciteSetBstMidEndSepPunct{\mcitedefaultmidpunct}
{\mcitedefaultendpunct}{\mcitedefaultseppunct}\relax
\EndOfBibitem
\bibitem{Adam:2015gba}
ALICE collaboration, J.~Adam {\em et~al.},
  \ifthenelse{\boolean{articletitles}}{\emph{{Measurement of an excess in the
  yield of \jpsi\ at very low $p_{\rm T}$ in Pb-Pb collisions at $\sqrt{s_{\rm
  NN}}$ = 2.76 TeV}},
  }{}\href{https://doi.org/10.1103/PhysRevLett.116.222301}{Phys.\ Rev.\ Lett.\
  \textbf{116} (2016) 222301},
  \href{http://arxiv.org/abs/1509.08802}{{\normalfont\ttfamily
  arXiv:1509.08802}}\relax
\mciteBstWouldAddEndPuncttrue
\mciteSetBstMidEndSepPunct{\mcitedefaultmidpunct}
{\mcitedefaultendpunct}{\mcitedefaultseppunct}\relax
\EndOfBibitem
\bibitem{PhysRevLett.123.132302}
STAR collaboration, Adam {\em et~al.},
  \ifthenelse{\boolean{articletitles}}{\emph{Observation of excess
  ${J}/\ensuremath{\psi}$ yield at very low transverse momenta in
  $\mathrm{Au}+\mathrm{Au}$ collisions at $\sqrt{{s}_{NN}}=200\text{ }\text{
  }\mathrm{GeV}$ and $\mathrm{U}+\mathrm{U}$ collisions at
  $\sqrt{{s}_{NN}}=193\text{ }\text{ }\mathrm{GeV}$},
  }{}\href{https://doi.org/10.1103/PhysRevLett.123.132302}{Phys.\ Rev.\ Lett.\
  \textbf{123} (2019) 132302},
  \href{http://arxiv.org/abs/1904.11658}{{\normalfont\ttfamily
  arXiv:1904.11658}}\relax
\mciteBstWouldAddEndPuncttrue
\mciteSetBstMidEndSepPunct{\mcitedefaultmidpunct}
{\mcitedefaultendpunct}{\mcitedefaultseppunct}\relax
\EndOfBibitem
\bibitem{PhysRev.45.729}
E.~J. Williams, \ifthenelse{\boolean{articletitles}}{\emph{Nature of the high
  energy particles of penetrating radiation and status of ionization and
  radiation formulae}, }{}\href{https://doi.org/10.1103/PhysRev.45.729}{Phys.\
  Rev.\  \textbf{45} (1934) 729}\relax
\mciteBstWouldAddEndPuncttrue
\mciteSetBstMidEndSepPunct{\mcitedefaultmidpunct}
{\mcitedefaultendpunct}{\mcitedefaultseppunct}\relax
\EndOfBibitem
\bibitem{BALTZ_2008}
A.~BALTZ {\em et~al.}, \ifthenelse{\boolean{articletitles}}{\emph{The physics
  of ultraperipheral collisions at the lhc},
  }{}\href{https://doi.org/10.1016/j.physrep.2007.12.001}{Physics Reports
  \textbf{458} (2008) 1–171},
  \href{http://arxiv.org/abs/0706.3356}{{\normalfont\ttfamily
  arXiv:0706.3356}}\relax
\mciteBstWouldAddEndPuncttrue
\mciteSetBstMidEndSepPunct{\mcitedefaultmidpunct}
{\mcitedefaultendpunct}{\mcitedefaultseppunct}\relax
\EndOfBibitem
\bibitem{ducati_heavy_2018}
M.~B.~G. Ducati and S.~Martins,
  \ifthenelse{\boolean{articletitles}}{\emph{Heavy meson photoproduction in
  peripheral {AA} collisions},
  }{}\href{https://doi.org/10.1103/PhysRevD.97.116013}{Phys.\ Rev.\
  \textbf{D97} (2018) 116013},
  \href{http://arxiv.org/abs/1804.09836}{{\normalfont\ttfamily
  arXiv:1804.09836}}\relax
\mciteBstWouldAddEndPuncttrue
\mciteSetBstMidEndSepPunct{\mcitedefaultmidpunct}
{\mcitedefaultendpunct}{\mcitedefaultseppunct}\relax
\EndOfBibitem
\bibitem{cepila}
Cepila, J. and Contreras, J. G. and Krelina, M.,
  \ifthenelse{\boolean{articletitles}}{\emph{Coherent and incoherent $J/\ensuremath{\psi}$ photonuclear production in an energy-dependent hot-spot model},
  }{}\href{https://link.aps.org/doi/10.1103/PhysRevC.97.024901}{Phys.\ Rev.\
  \textbf{C97} (2018) 024901},
  \href{http://arxiv.org/abs/1711.01855}{{\normalfont\ttfamily
  arXiv:1711.01855}}\relax
\mciteBstWouldAddEndPuncttrue
\mciteSetBstMidEndSepPunct{\mcitedefaultmidpunct}
{\mcitedefaultendpunct}{\mcitedefaultseppunct}\relax
\EndOfBibitem
\bibitem{PhysRevC.97.044910}
W.~Zha {\em et~al.}, \ifthenelse{\boolean{articletitles}}{\emph{Coherent
  ${J}/\ensuremath{\psi}$ photoproduction in hadronic heavy-ion collisions},
  }{}\href{https://doi.org/10.1103/PhysRevC.97.044910}{Phys.\ Rev.\
  \textbf{C97} (2018) 044910},
  \href{http://arxiv.org/abs/1705.01460}{{\normalfont\ttfamily
  arXiv:1705.01460}}\relax
\mciteBstWouldAddEndPuncttrue
\mciteSetBstMidEndSepPunct{\mcitedefaultmidpunct}
{\mcitedefaultendpunct}{\mcitedefaultseppunct}\relax
\EndOfBibitem
\bibitem{LHCb-DP-2008-001}
LHCb collaboration, A.~A. Alves~Jr.\ {\em et~al.},
  \ifthenelse{\boolean{articletitles}}{\emph{{The \lhcb detector at the LHC}},
  }{}\href{https://doi.org/10.1088/1748-0221/3/08/S08005}{JINST \textbf{3}
  (2008) S08005}\relax
\mciteBstWouldAddEndPuncttrue
\mciteSetBstMidEndSepPunct{\mcitedefaultmidpunct}
{\mcitedefaultendpunct}{\mcitedefaultseppunct}\relax
\EndOfBibitem
\bibitem{LHCb-DP-2014-002}
LHCb collaboration, R.~Aaij {\em et~al.},
  \ifthenelse{\boolean{articletitles}}{\emph{{LHCb detector performance}},
  }{}\href{https://doi.org/10.1142/S0217751X15300227}{Int.\ J.\ Mod.\ Phys.\
  \textbf{A30} (2015) 1530022},
  \href{http://arxiv.org/abs/1412.6352}{{\normalfont\ttfamily
  arXiv:1412.6352}}\relax
\mciteBstWouldAddEndPuncttrue
\mciteSetBstMidEndSepPunct{\mcitedefaultmidpunct}
{\mcitedefaultendpunct}{\mcitedefaultseppunct}\relax
\EndOfBibitem
\bibitem{LHCb-DP-2014-001}
LHCb collaboration, R.~Aaij {\em et~al.},
  \ifthenelse{\boolean{articletitles}}{\emph{{Performance of the LHCb Vertex Locator}},
  }{}\href{https://doi.org/10.1088/1748-0221/9/09/P09007}{JINST
  \textbf{9} (2014) P09007},
  \href{http://arxiv.org/abs/1405.7808}{{\normalfont\ttfamily
  arXiv:1405.7808}}\relax
\mciteBstWouldAddEndPuncttrue
\mciteSetBstMidEndSepPunct{\mcitedefaultmidpunct}
{\mcitedefaultendpunct}{\mcitedefaultseppunct}\relax
\EndOfBibitem
\bibitem{LHCb-DP-2017-001}
d'Argent, Ph. and others,
  \ifthenelse{\boolean{articletitles}}{\emph{{Improved performance of the LHCb Outer Tracker in LHC Run 2}},
  }{}\href{https://doi.org/10.1088/1748-0221/12/11/P11016}{JINST
  \textbf{12} (2017) P11016},
  \href{http://arxiv.org/abs/1708.00819}{{\normalfont\ttfamily
  arXiv:1708.00819}}\relax
\mciteBstWouldAddEndPuncttrue
\mciteSetBstMidEndSepPunct{\mcitedefaultmidpunct}
{\mcitedefaultendpunct}{\mcitedefaultseppunct}\relax
\EndOfBibitem
\bibitem{LHCb-DP-2012-003}
Adinolfi, M. and others,
  \ifthenelse{\boolean{articletitles}}{\emph{{Performance of the \lhcb RICH detector at the LHC}},
  }{}\href{https://doi.org/10.1140/epjc/s10052-013-2431-9}{Eur. Phys. J.
  \textbf{C73} (2013) 2431},
  \href{http://arxiv.org/abs/1211.6759}{{\normalfont\ttfamily
  arXiv:1211.6759}}\relax
\mciteBstWouldAddEndPuncttrue
\mciteSetBstMidEndSepPunct{\mcitedefaultmidpunct}
{\mcitedefaultendpunct}{\mcitedefaultseppunct}\relax
\EndOfBibitem
\bibitem{LHCb-DP-2012-002}
Alves Jr., A A and others,
  \ifthenelse{\boolean{articletitles}}{\emph{{Performance of the LHCb muon system}},
  }{}\href{https://doi.org/10.1088/1748-0221/8/02/P02022}{JINST
  \textbf{8} (2013) P02022},
  \href{http://arxiv.org/abs/1211.1346}{{\normalfont\ttfamily
  arXiv:1211.1346}}\relax
\mciteBstWouldAddEndPuncttrue
\mciteSetBstMidEndSepPunct{\mcitedefaultmidpunct}
{\mcitedefaultendpunct}{\mcitedefaultseppunct}\relax
\EndOfBibitem
\bibitem{10.1093/ptep/ptaa104}
Particle Data Group, P.~A.~{\relax Zyla}. et~al.\,
  \ifthenelse{\boolean{articletitles}}{\emph{{Review of Particle Physics}},
  }{}\href{https://doi.org/10.1093/ptep/ptaa104}{Progress of Theoretical and
  Experimental Physics \textbf{2020} (2020) }, 083C01\relax
\mciteBstWouldAddEndPuncttrue
\mciteSetBstMidEndSepPunct{\mcitedefaultmidpunct}
{\mcitedefaultendpunct}{\mcitedefaultseppunct}\relax
\EndOfBibitem
\bibitem{LHCb-PAPER-2014-047}
LHCb collaboration, R.~Aaij {\em et~al.},
  \ifthenelse{\boolean{articletitles}}{\emph{{Precision luminosity measurements
  at LHCb}}, }{}\href{https://doi.org/10.1088/1748-0221/9/12/P12005}{JINST
  \textbf{9} (2014) P12005},
  \href{http://arxiv.org/abs/1410.0149}{{\normalfont\ttfamily
  arXiv:1410.0149}}\relax
\mciteBstWouldAddEndPuncttrue
\mciteSetBstMidEndSepPunct{\mcitedefaultmidpunct}
{\mcitedefaultendpunct}{\mcitedefaultseppunct}\relax
\EndOfBibitem
\bibitem{PhysRevC.97.054910}
C.~Loizides, J.~Kamin, and D.~d'Enterria,
  \ifthenelse{\boolean{articletitles}}{\emph{Improved {M}onte {C}arlo {G}lauber
  predictions at present and future nuclear colliders},
  }{}\href{https://doi.org/10.1103/PhysRevC.97.054910}{Phys.\ Rev.\
  \textbf{C97} (2018) 054910},
  \href{http://arxiv.org/abs/1710.07098}{{\normalfont\ttfamily
  arXiv:1710.07098}}\relax
\mciteBstWouldAddEndPuncttrue
\mciteSetBstMidEndSepPunct{\mcitedefaultmidpunct}
{\mcitedefaultendpunct}{\mcitedefaultseppunct}\relax
\EndOfBibitem
\bibitem{Abelev:2013qoq}
{ALICE collaboration}, \ifthenelse{\boolean{articletitles}}{\emph{Centrality
  determination of {Pb}-{Pb} collisions at {$\sqrt{s_{\rm{NN}}}=2.76$\tev} with
  {ALICE}}, }{}\href{https://doi.org/10.1103/PhysRevC.88.044909}{Phys.\ Rev.\
  \textbf{C88} (2013) 044909},
  \href{http://arxiv.org/abs/1301.4361}{{\normalfont\ttfamily
  arXiv:1301.4361}}\relax
\mciteBstWouldAddEndPuncttrue
\mciteSetBstMidEndSepPunct{\mcitedefaultmidpunct}
{\mcitedefaultendpunct}{\mcitedefaultseppunct}\relax
\EndOfBibitem
\bibitem{LHCb-DP-2021-001}
R.~Aaij {\em et~al.}, \ifthenelse{\boolean{articletitles}}{\emph{{Centrality
  determination in heavy-ion collisions with the LHCb detector}}, }{}
  LHCb-DP-2021-001, {in preparation}\relax
\mciteBstWouldAddEndPuncttrue
\mciteSetBstMidEndSepPunct{\mcitedefaultmidpunct}
{\mcitedefaultendpunct}{\mcitedefaultseppunct}\relax
\EndOfBibitem

\bibitem{supp_mat}
See Supplemental Material at [URL will be inserted by publisher] for additional information on the centrality measurement and detailed results of the photo-produced \jpsi\ mesons measurement.
\EndOfBibitem

\bibitem{Pivk:2004ty}
M.~Pivk and F.~R. Le~Diberder,
  \ifthenelse{\boolean{articletitles}}{\emph{{\sPlot: A statistical tool to
  unfold data distributions}},
  }{}\href{https://doi.org/10.1016/j.nima.2005.08.106}{Nucl.\ Instrum.\ Meth.\
  \textbf{A555} (2005) 356},
  \href{http://arxiv.org/abs/physics/0402083}{{\normalfont\ttfamily
  arXiv:physics/0402083}}\relax
\mciteBstWouldAddEndPuncttrue
\mciteSetBstMidEndSepPunct{\mcitedefaultmidpunct}
{\mcitedefaultendpunct}{\mcitedefaultseppunct}\relax
\EndOfBibitem
\bibitem{Skwarnicki:1986xj}
T.~Skwarnicki, {\em {A study of the radiative cascade transitions between the
  Upsilon-prime and Upsilon resonances}}, PhD thesis, Institute of Nuclear
  Physics, Krakow, 1986,
  {\href{http://inspirehep.net/record/230779/}{DESY-F31-86-02}}\relax
\mciteBstWouldAddEndPuncttrue
\mciteSetBstMidEndSepPunct{\mcitedefaultmidpunct}
{\mcitedefaultendpunct}{\mcitedefaultseppunct}\relax
\EndOfBibitem
\bibitem{EPOS}
T.~Pierog {\em et~al.}, \ifthenelse{\boolean{articletitles}}{\emph{{EPOS LHC:
  Test of collective hadronization with data measured at the CERN Large Hadron
  Collider}}, }{}\href{https://doi.org/10.1103/PhysRevC.92.034906}{Phys.\ Rev.\
   \textbf{C92} (2015) 034906},
  \href{http://arxiv.org/abs/1306.0121}{{\normalfont\ttfamily
  arXiv:1306.0121}}\relax
\mciteBstWouldAddEndPuncttrue
\mciteSetBstMidEndSepPunct{\mcitedefaultmidpunct}
{\mcitedefaultendpunct}{\mcitedefaultseppunct}\relax
\EndOfBibitem
\bibitem{Sjostrand:2007gs}
T.~Sj\"{o}strand, S.~Mrenna, and P.~Skands,
  \ifthenelse{\boolean{articletitles}}{\emph{{A brief introduction to PYTHIA
  8.1}}, }{}\href{https://doi.org/10.1016/j.cpc.2008.01.036}{Comput.\ Phys.\
  Commun.\  \textbf{178} (2008) 852},
  \href{http://arxiv.org/abs/0710.3820}{{\normalfont\ttfamily
  arXiv:0710.3820}}\relax
\mciteBstWouldAddEndPuncttrue
\mciteSetBstMidEndSepPunct{\mcitedefaultmidpunct}
{\mcitedefaultendpunct}{\mcitedefaultseppunct}\relax
\EndOfBibitem
\bibitem{Sjostrand:2006za}
T.~Sj\"{o}strand, S.~Mrenna, and P.~Skands,
  \ifthenelse{\boolean{articletitles}}{\emph{{PYTHIA 6.4 physics and manual}},
  }{}\href{https://doi.org/10.1088/1126-6708/2006/05/026}{JHEP \textbf{05}
  (2006) 026}, \href{http://arxiv.org/abs/hep-ph/0603175}{{\normalfont\ttfamily
  arXiv:hep-ph/0603175}}\relax
\mciteBstWouldAddEndPuncttrue
\mciteSetBstMidEndSepPunct{\mcitedefaultmidpunct}
{\mcitedefaultendpunct}{\mcitedefaultseppunct}\relax
\EndOfBibitem
\bibitem{LHCb-PROC-2010-056}
I.~Belyaev {\em et~al.}, \ifthenelse{\boolean{articletitles}}{\emph{{Handling
  of the generation of primary events in Gauss, the LHCb simulation
  framework}}, }{}\href{https://doi.org/10.1088/1742-6596/331/3/032047}{J.\
  Phys.\ Conf.\ Ser.\  \textbf{331} (2011) 032047}\relax
\mciteBstWouldAddEndPuncttrue
\mciteSetBstMidEndSepPunct{\mcitedefaultmidpunct}
{\mcitedefaultendpunct}{\mcitedefaultseppunct}\relax
\EndOfBibitem
\bibitem{starlight}
S.~Klein {\em et~al.}, \ifthenelse{\boolean{articletitles}}{\emph{{STAR}light:
  {A} {M}onte {C}arlo simulation program for ultra-peripheral collisions of
  relativistic ions},
  }{}\href{https://doi.org/10.1016/j.cpc.2016.10.016}{Computer Physics
  Communications \textbf{212} (2017) 258},
  \href{http://arxiv.org/abs/1607.03838}{{\normalfont\ttfamily
  arXiv:1607.03838}}\relax
\mciteBstWouldAddEndPuncttrue
\mciteSetBstMidEndSepPunct{\mcitedefaultmidpunct}
{\mcitedefaultendpunct}{\mcitedefaultseppunct}\relax
\EndOfBibitem
\bibitem{Allison:2006ve}
Geant4 collaboration, J.~Allison {\em et~al.},
  \ifthenelse{\boolean{articletitles}}{\emph{{Geant4 developments and
  applications}}, }{}\href{https://doi.org/10.1109/TNS.2006.869826}{IEEE
  Trans.\ Nucl.\ Sci.\  \textbf{53} (2006) 270}\relax
\mciteBstWouldAddEndPuncttrue
\mciteSetBstMidEndSepPunct{\mcitedefaultmidpunct}
{\mcitedefaultendpunct}{\mcitedefaultseppunct}\relax
\EndOfBibitem
\bibitem{Agostinelli:2002hh}
Geant4 collaboration, S.~Agostinelli {\em et~al.},
  \ifthenelse{\boolean{articletitles}}{\emph{{Geant4: A simulation toolkit}},
  }{}\href{https://doi.org/10.1016/S0168-9002(03)01368-8}{Nucl.\ Instrum.\
  Meth.\  \textbf{A506} (2003) 250}\relax
\mciteBstWouldAddEndPuncttrue
\mciteSetBstMidEndSepPunct{\mcitedefaultmidpunct}
{\mcitedefaultendpunct}{\mcitedefaultseppunct}\relax
\EndOfBibitem
\bibitem{LHCb-PROC-2011-006}
M.~Clemencic {\em et~al.}, \ifthenelse{\boolean{articletitles}}{\emph{{The
  \lhcb simulation application, Gauss: Design, evolution and experience}},
  }{}\href{https://doi.org/10.1088/1742-6596/331/3/032023}{J.\ Phys.\ Conf.\
  Ser.\  \textbf{331} (2011) 032023}\relax
\mciteBstWouldAddEndPuncttrue
\mciteSetBstMidEndSepPunct{\mcitedefaultmidpunct}
{\mcitedefaultendpunct}{\mcitedefaultseppunct}\relax
\EndOfBibitem
\bibitem{tsallis_nonextensive_1999}
C.~Tsallis, \ifthenelse{\boolean{articletitles}}{\emph{Nonextensive statistics:
  theoretical, experimental and computational evidences and connections},
  }{}\href{https://doi.org/10.1590/S0103-97331999000100002}{Brazilian Journal
  of Physics \textbf{29} (1999) 1}\relax
\mciteBstWouldAddEndPuncttrue
\mciteSetBstMidEndSepPunct{\mcitedefaultmidpunct}
{\mcitedefaultendpunct}{\mcitedefaultseppunct}\relax
\EndOfBibitem
\bibitem{PhysRevC.99.061901}
W.~Zha, L.~Ruan, Z.~Tang, Z.~Xu, S.~Yang, \ifthenelse{\boolean{articletitles}}{\emph{Double-slit
  experiment at {F}ermi scale: Coherent photoproduction in heavy-ion
  collisions}, }{}\href{https://doi.org/10.1103/PhysRevC.99.061901}{Phys.\
  Rev.\  \textbf{C99} (2019) 061901},
  \href{http://arxiv.org/abs/1810.10694}{{\normalfont\ttfamily
  arXiv:1810.10694}}\relax
\mciteBstWouldAddEndPuncttrue
\mciteSetBstMidEndSepPunct{\mcitedefaultmidpunct}
{\mcitedefaultendpunct}{\mcitedefaultseppunct}\relax
\EndOfBibitem
\end{mcitethebibliography}
\mciteErrorOnUnknownfalse

\ifx\mcitethebibliography\mciteundefinedmacro
\PackageError{LHCb.bst}{mciteplus.sty has not been loaded}
{This bibstyle requires the use of the mciteplus package.}\fi
\providecommand{\href}[2]{#2}

\newpage

\centerline
{\large\bf LHCb collaboration}
\begin
{flushleft}
\small
R.~Aaij$^{32}$,
C.~Abell{\'a}n~Beteta$^{50}$,
T.~Ackernley$^{60}$,
B.~Adeva$^{46}$,
M.~Adinolfi$^{54}$,
H.~Afsharnia$^{9}$,
C.A.~Aidala$^{85}$,
S.~Aiola$^{25}$,
Z.~Ajaltouni$^{9}$,
S.~Akar$^{65}$
% ...

{\footnotesize \it

% ...
$^{9}$Universit{\'e} Clermont Auvergne, CNRS/IN2P3, LPC, Clermont-Ferrand, France\\
% ...
$^{25}$INFN Sezione di Milano, Milano, Italy\\
% ...
$^{32}$Nikhef National Institute for Subatomic Physics, Amsterdam, Netherlands\\
% ...
$^{46}$Instituto Galego de F{\'\i}sica de Altas Enerx{\'\i}as (IGFAE), Universidade de Santiago de Compostela, Santiago de Compostela, Spain\\
% ...
$^{50}$Physik-Institut, Universit{\"a}t Z{\"u}rich, Z{\"u}rich, Switzerland\\
% ...
$^{54}$H.H. Wills Physics Laboratory, University of Bristol, Bristol, United Kingdom\\
% ...
$^{60}$Oliver Lodge Laboratory, University of Liverpool, Liverpool, United Kingdom\\
% ...
$^{65}$University of Cincinnati, Cincinnati, OH, United States\\
% ...
$^{68}$Syracuse University, Syracuse, NY, United States\\
% ...
$^{85}$University of Michigan, Ann Arbor, United States, associated to $^{68}$\\
\bigskip
% ...
\medskip
}
\end{flushleft}

\end{document}